\documentclass[a4paper,12pt]{article}
\usepackage{graphicx}
\usepackage{authblk}
\usepackage{cite}
\usepackage[english]{babel}
\usepackage{setspace}
\usepackage{relsize}
\usepackage{bm} 
\title{Remarks on energetic conditions for positronium formation in non-polar solids. Coupled Dipole Method application.}
\author[]{M.~Pietrow\thanks{email: mrk@kft.umcs.lublin.pl}}
\affil[]{Institute of Physics, M. Curie-Sk{\l}odowska University, ul.~Pl.~M.~Curie-Sk{\l}odowskiej~1, 20-031 Lublin, Poland}
\begin{document}
\maketitle
\section{Abstract}
A numerical program calculating an energy of a positron or (and) an electron near the free volume in solid n-alkanes has been build. The theory of interaction of $e^+$ or (and) $e^-$ with this non-polar media based on polarizability has been introduced. The energy of the $e^+$ -- $e^-$ pair in the bulk was compared to that calculated when the pair forms a positronium (Ps) inside the free volume. The calculations are based on the Coupled Dipole Method and the dipole-dipole interaction energy for induced dipoles is taken into account. Furthermore, a correction of a local permittivity for the $e^+$--$e^-$ interaction is calculated taking into account the non-isotropic medium between them. The method is a step toward more accurate calculations of energetic conditions during the Ps formation in matter.\\
The possibility of emission of the excess energy of the Ps formation as electromagnetic radiation is discussed. It is argued that if this radiation is observed, it can be used as a new spectroscopic tool providing information about microscopic properties of media.
\paragraph{Keywords:} positronium formation, electron traps, crystals of alkanes, polarizability
\paragraph{PACS:} 36.10.Dr, 78.70.Bj, 31.10.+z
\section{Introduction}
The interaction of positron with matter is a subject of radiation physics and chemistry \cite{Mozumder,Stepanov03}. However, in this case, the energetic positron is subjected. At this scale of energy, the interactions resemble these for an electron~--~both particles can be classified as light charged particles. When these particles are quasi-thermalized, the timescale of the interaction is nanoseconds and chemical type of phenomena as spur formation or solvatation should be take into account \cite{MozumderHatano}. Although a chunk of the published theory on thermalized electron in matter could be applied to the positron, the theory of electron solvatation and its mobility in matter cannot be applied to the positron directly. Among the differences, one could list a quantum effect of the exchange interaction with molecule electrons. Furthermore, some other features such as the timescale of relaxation, due to a positron annihilation, cannot be transferred. These difficulties cause that there is not present in the literature a quantum model of positron interaction with condensed matter. Furthermore, since the problem of positronium formation in matter is related to many-body interactions of the electron and positron pair with the bulk, only an approximate description of the Ps formation (classical or quasi-classical) is used at present. As a consequence, experimental techniques based on positron annihilation (e.g. Positron Annihilation Lifetime Spectroscopy~\cite{Goworek14}) require still more accurate models of positronium formation to interpret their results credibly.\\
Among discussed questions, energetic conditions during the Ps formation are the crucial ones. According to the model~\cite{Stepanov03}, a positron from a $\beta^{+}$ radiation source produces ions and quasi-free electrons during its slow-down in matter. When the kinetic energy of the positron falls below the ionization energy level, the rest of the energy is transformed into electronic orbital excitations and vibrations of a lattice. Finally, a positron can annihilate with one of the electrons from the bulk (free annihilation) or can form an intermediate (unstable) state of a positronium atom and then annihilate either with the electron from this atom or with an electron from the bulk (pick-off).\\
Some conditions should be fulfilled for Ps to be formed. One of the conditions is the presence of a free volume of the radius sufficient to store the atom of the radius $3 a_0\cdot n^2$, where $a_0\simeq$0.53$\AA$ (approximated value taken from Bohr's model of the atom in the vacuum)\footnote{In fact, the better approximation of the minimum radius of the free volume originates from the quantum model of Ps (regarded as an undivided quantum particle in spherical quantum well) \cite{Consolati14}. This radius is approximated as 1.66~$\AA$ in this case.}.\\
Not of less importance for the theory of positronium formation is to consider the problem of energetic conditions~-~before and after Ps creation. Results of these considerations have a direct application in the interpretation of results generated from positron-based experimental techniques.\\
Energetic conditions in Ps formation are considered in \cite{Stepanov13}. There, the change in internal energy during Ps formation is estimated generally for a wide range of materials by the difference in the interaction energy of $e^+$ and $e^-$ when the relative permittivity of a medium between these particles is $\epsilon_r$ and when $\epsilon_r$=1 in the free volume when Ps is created. It was noted that the quantum model of the interaction of $e^+$ and $e^-$ with the medium requires estimation of the value of particle-medium interaction terms for each of these particles in a particular case. These values were approximated as a work function and taken from the experimental data tables for known materials.\\
Here, we try to make more accurate and systematic numerical calculations of the energetic conditions for n-alkanes. Alkanes are an example of non-polar organic medium. Although their molecules do not have a permanent dipole moment, alkanes have relatively great values of polarizability \cite{crc_Handbook} (a ratio of an induced dipole moment to the electric field that induces this moment). Therefore, the induced dipole moments allowed us to explain the nature of electron traps and their depth in alkanes \cite{Pietrow15}. Now, we try to generalize the calculations based on polarizability and calculate the coupling energy of $e^+$ or (and) $e^-$ to the surroundings not only in the case of trapping. The calculations described here are based on the Coupled Dipole Method (CDM) describing the potential energy of induced dipoles in an electrostatic field \cite{Kwaadgras14,Kwaadgras14b}. It seems that the application of the CDM method allows 'microscopic' analysis of the problem of profitability to enter the free volume by the $e^+$--$e^-$~-~pair and formation of Ps. A future study based on these calculations will provide an answer to the following questions: what does it mean that $e^-$ drags behind $e^+$ when escaping a blob region and forms a so~-~called quasi-free Ps \cite{Stepanov03} (why, say, is qf-Ps$^-$ not preferred?)? What is the energetic difference between the formation of the Ps with the use of a blob electron or a trapped electron?\\
In the next part of the paper, the symbol $e^=$ is used when the subject of discussion refers equally well to the positron $e^+$ and the electron $e^-$.
\section{Description of the model}\label{Description of the model}
We assume that the main part of the energy of interaction of $e^=$ with the molecules of the medium can be approximated by the energy of induced dipoles in an electric field that induces them. In our case, the source of this field is $e^=$ located among particle chains. The interesting model of these interactions, including the dipole-dipole interaction, was introduced by Kwaadgras et al. \cite{Kwaadgras14}. The model is based on the notion of Lorentz atom (a piece of matter with an inducible dipole moment). Each molecule is divided into Lorentz atoms called \emph{chunks}.\\
The CDM Hamiltonian for $N$ chunks in the system is
\begin{equation}
H=\frac{1}{2}\sum_{i=1}^{N}\mathbf{k}_i\cdot \mathbf{m}_i^{-1}\cdot \mathbf{k}_i+\frac{1}{2}\sum_{i,j=1}^{N} \mathbf{d}_i\cdot\mathbf{q}_i\cdot(\bm{\alpha}_i^{-1}\delta_{ij}-\mathbf{T}_{ij})\cdot \mathbf{q}_j \cdot \mathbf{d}_j-\sum_{i=1}^{N}(\mathbf{q}_i \cdot\mathbf{d}_i)\cdot \mathbf{E}_0^{(i)}
\end{equation}
where
\begin{equation}
\mathbf{T}_{ij}=\left\{ \begin{array}{ccl}
(\frac{3\ |\mathbf{r}_{ij}\rangle\langle\mathbf{r}_{ij}|}{r^2_{ij}}-\mathbf{I})/r^3_{ij} & \mbox{for} & i\neq j \\
\mathbf{0} & \mbox{for} & i=j
\end{array}\right.
\end{equation}
is a dipole tensor and $\mathbf{m}_i\equiv m_i\mathbf{I}$, $\mathbf{q}_i\equiv q_i\mathbf{I}$ are $3N\times 3N$-dimensional matrices consisting of masses $m_i$ and charges $q_i$, respectively, for the $i$-th Lorentz atom, whereas $\mathbf{I}$ is the 3$\times$3 identity matrix; $\mathbf{k}_i$ is a momentum vector related to the $i$-th Lorentz atom, $\mathbf{d}_i$~--~charge separation distance in it, $\bm{\alpha}_i$~--~polarizability tensor, and $\mathbf{E}_0^{(i)}$ is an electric field in place of the $i$-th chunk.\\
Some transformations described in \cite{Kwaadgras14} allow making this Hamiltonian resemble that for a set of oscillators. Finally, the energy of the interaction of induced dipoles with an electric field (with $e^+$ or $e^-$, in our case) can be separated. It reads as
\begin{equation}
U_E=-\frac{1}{2}\mathcal{P}\cdot \mathlarger{\mathlarger{\epsilon}}_0,
\end{equation}
where the effective dipole moment $\mathcal{P}$ can be calculated from $(\mathcal{A}^{-1}-\mathcal{T})^{-1}\cdot \mathcal{P}=\mathlarger{\mathlarger{\mathlarger{\epsilon}}}_0$, where $\mathcal{A}\equiv~diag(\{\bm{\alpha}_i\})$, $\mathlarger{\mathlarger{\mathlarger{\epsilon}}}_0\equiv diag(\{\mathbf{E}_0^{(i)}\})$. $\mathcal{T}$ is constructed as follows from $\bm{T}_{ij}$ in the way given as an example for a case of two chunks,
\begin{equation}
\mathcal{T} = \left(
\begin{array}{ccc|ccc}
0 & 0 & 0 & T_{1,2}^{x,x} & T_{1,2}^{x,y} & T_{1,2}^{x,z} \\
0 & 0 & 0 & T_{1,2}^{y,x} & T_{1,2}^{y,y} & T_{1,2}^{y,z} \\
0 & 0 & 0 & T_{1,2}^{z,x} & T_{1,2}^{z,y} & T_{1,2}^{z,z} \\ \hline
T_{2,1}^{x,x} & T_{2,1}^{x,y} & T_{2,1}^{x,z} & 0 & 0 & 0 \\
T_{2,1}^{y,x} & T_{2,1}^{y,y} & T_{2,1}^{y,z} & 0 & 0 & 0 \\
T_{2,1}^{z,x} & T_{2,1}^{z,y} & T_{2,1}^{z,z} & 0 & 0 & 0
\end{array} \right).
\end{equation}
The non-trivial part of the analysis of positron (electron) interaction near the free volume is also the mutual interaction of these particles. In the scale of a free volume size (tenths of $\AA$) we could not assume that the environment between them is homogeneous. $e^+$ and $e^-$ are separated by only some elongated molecules. The induced dipole moment components along and perpendicularly to the molecules do differ considerably.\\
The approach made by us to estimate the effective permittivity in this case is based on a model of inclusions of axial particles of permittivity $\epsilon_1$ immersed in the environment of the permittivity $\epsilon_2$. Here, we assumed that $\epsilon_1=\epsilon_r \epsilon_0\simeq 2\epsilon_0$ \cite{crc_Handbook}, and $\epsilon_2=\epsilon_0$, where $\epsilon_0$ is vacuum permittivity. Applying the formula
\begin{equation}
\epsilon_i=\epsilon_0+\frac{n \alpha_i}{1-L_i \frac{n \alpha_i}{\epsilon_0}}
\end{equation}
given in \cite{Sareni}, we obtain the relative permittivity for $x$ ($y$) and $z$ direction respectively: $\epsilon_x^{(r)}=\epsilon_y^{(r)}$=1.33, $\epsilon_z^{(r)}$=2.71. Here we assumed $\alpha_x=\alpha_y=\alpha_z/4=\alpha/4$, where the $\alpha$ value is cited below. $n=1/(l\cdot w\cdot h)$, where $l$, $w$, $h$~--~lengths of the particle along $x$, $y$, $z$ dimensions. $L_i$ are so~-~called depolarization factors \cite{Sareni} in the direction $i$. In particular, $L_x$ can be calculated as
\begin{equation}
L_x=\frac{abc}{2}\int_0^{\infty}\frac{du}{(u+a^2)\sqrt{(u+a^2)(u+b^2)(u+c^2)}},
\end{equation}
where $a$, $b$, $c$ are semi-axes of an ellipsoid representing the particle. $L_y$ and $L_z$ can be calculated after interchange of $a$, $b$, $c$\footnote{We should note that another way of calculation of the relative permittivity, using a formula
\begin{equation}
\epsilon_i=\epsilon_0 \left( 1+\frac{(\epsilon_1-\epsilon_0)f}{\epsilon_0+(\epsilon_1-\epsilon_0)(1-f)L_i} \right)
\end{equation}
from this paper, requires calculation of the volume fraction $f$ of a cylindrical body in the volume cell. After calculating this and inserting $\epsilon_1=\epsilon_r\epsilon_0$, where $\epsilon_r\sim$2, we have $\epsilon_x^{(r)}=\epsilon_y^{(r)}$=1.61, whereas $\epsilon_z^{(r)}$=1.44. There is a considerable difference in the results of both ways of calculation.}.
\\
With the support of the formula \cite{Greiner}
\begin{equation}
U_{ee}=-\frac{ke^2}{\sqrt{det \hat{\epsilon}}\quad \sqrt{\frac{x_{ee}^2}{\epsilon_x}+\frac{y_{ee}^2}{\epsilon_y}+\frac{z_{ee}^2}{\epsilon_z}}},
\end{equation}
where $x_{ee}$, $y_{ee}$, $z_{ee}$ are distances between electron and positron in $x$, $y$, $z$ directions, whereas $k=1/(4\pi\epsilon_0)$ and
\begin{displaymath}
\hat{\epsilon} =
\left( \begin{array}{ccc}
\epsilon_x & 0 & 0 \\
0 & \epsilon_y & 0 \\
0 & 0 & \epsilon_z
\end{array} \right),
\end{displaymath}
we can calculate the energy of interaction for them in the presence of an anisotropic medium under study.
\\
%
\section{Numerical assumptions}\label{Numerical assumptions}
In numerical calculations, we assumed that a chunk is equivalent to one segment --CH$_2$-- or CH$_3$-- of alkane chain. In particular, the calculations were performed for docosane (C$_{22}$H$_{46}$), thus each molecule consisted of $n$=22 chunks. The translational momenta of every chunks were assumed equal to 0.
\\
The size of the molecules was assumed as 7~$\AA$ $\times$ 4~$\AA$ $\times$ 30~$\AA$ \cite{Wentzel}. The gap between layers of molecules was set at 5~$\AA$.
\\
The polarizability for docosane was taken as extrapolation of data from \cite{crc_Handbook} given for shorter chains; it equals $\alpha$=40$\times$10$^{-24}$~cm$^3$[CGS]=4.45038$\times$10$^{-39}$~F$\cdot$m$^2$[SI]. The polarizability tensor of a chunk was calculated in the following way: we assumed that chunks obeyed cylindrical symmetry, so $\alpha_x=\alpha_y$ for each chunk. We also assumed $\alpha_x=\eta\alpha_z$, where $\eta$=1/4 as suggested in \cite{Kwaadgras14} for cylindrical molecules described there. Finally, we also assumed that $\alpha_z=\alpha/n$.\\
For molecules which form non-planar conformers (\emph{kink}, \emph{gauche} conformation), a quadratic B\'{e}zier curve was built based on coordinations of the chunk and those of its nearest neighbors. Then, the normed tangent vector for this curve 
is calculated in the middle point (representing the analyzed chunk). The ratio of the components of this vector allows setting the weights of partition of the polarizability $\{\alpha^x,\alpha^y,\alpha^z\}$ for this chunk.
\\
The idea of the program (written in \emph{Mathematica} software by Wolfram Research) is as follows. Firstly, a chunk (the smallest segment of a molecule) temporarily representing a molecule is chosen according to the position of the molecule in a molecular crystal lattice. A hexagonal lattice was assumed for the calculations. Although it is not the only crystallographic system observed in alkanes \cite{Turner}, the lattice type seems not to be a key parameter in our considerations. The free volume in the sample was generated as a vacancy defect~--~one or more molecules were absent in their places.\\
After generation of each initial chunk, next chunks started to grow. Planar conformers (\emph{all-trans}) were built by the consecutive grow of the chunks in the molecule along the $z$-axis of the crystal. Non~-~planar conformers were generated by choosing of the position of the next chunk not along the \emph{z}-axis of the crystal but in some distorted position (the distance between chunks was kept). The probability of the distortion is set by the numerical parameter.\\
The number of generated molecules is determined by the interaction range of e$^=$. This range is defined as the radius of a sphere embracing molecules, centered in place of e$^=$, for which further spreading of this region does not change the energy of electrostatic interactions by more than 0.01~eV. In this empirical way, the range of $e^=$ interaction was set at 40~$\AA$.
\section{Results}
\begin{figure}
\centering
\includegraphics[scale=0.25]{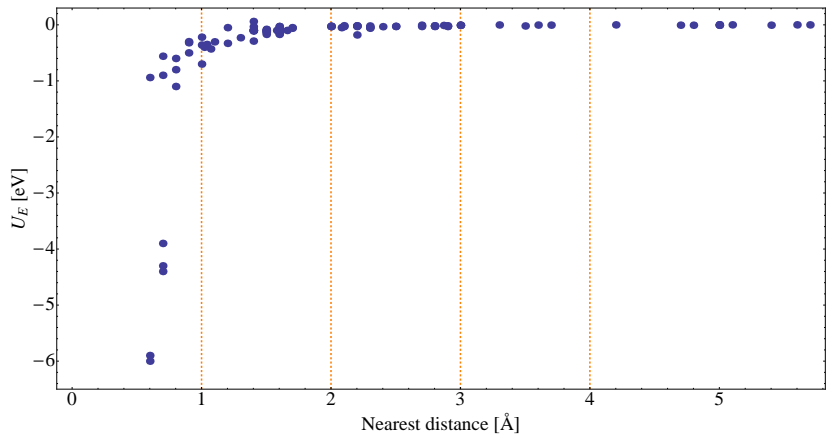}
\caption{$U_E$ as a function of the nearest chunk distance. One point in the figure is related to one random choice of the electron position in the volume.}
\label{fig:UE(r)}
\end{figure}

\begin{figure}
\centering
\includegraphics[scale=0.25]{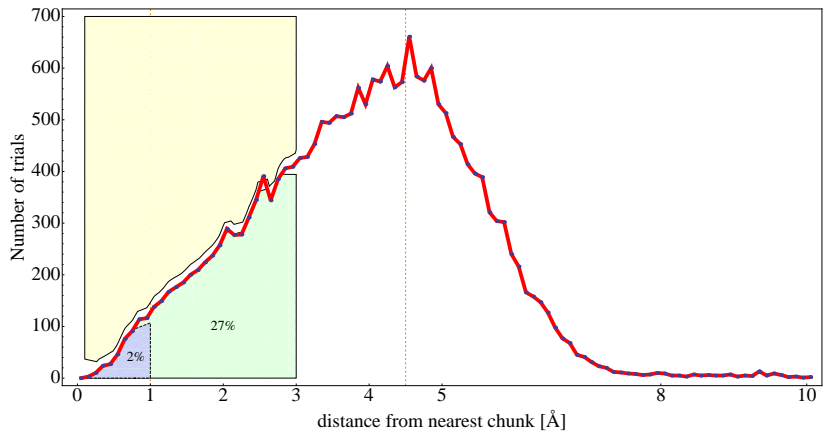}
\caption{The number of cases of settling a given distance from the nearest chunk. The electron positions were set randomly inside the sample. The whole number of trials was 20~thousand. Two regions are marked: the percentage of the population that lands closer than 3~$\AA$ and that closer than~1~$\AA$.}
\label{fig:RandomPlace}
\end{figure}

\begin{figure}
\centering
\includegraphics[scale=0.5]{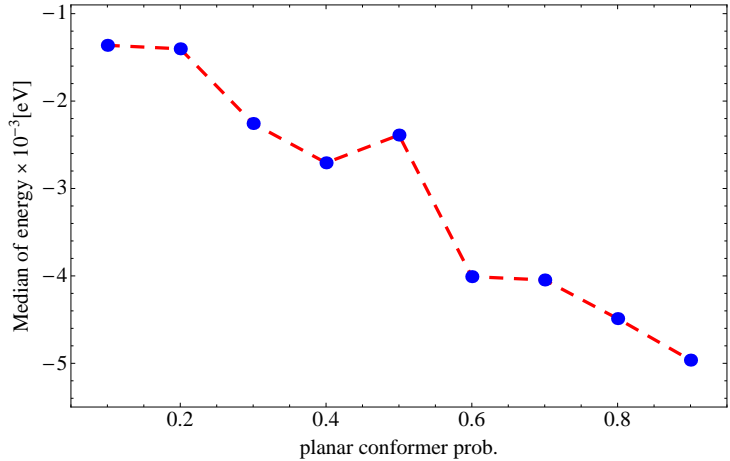}
\caption{Median value of $U_E$ for one electron inside a sample calculated for a medium with different fractions of non-planar conformers. The values on the abscissa denote a probability of forming the planar conformer during the procedure of 'growth' of the sample (see sec.~\ref{Numerical assumptions}). Each point is a median value of 300 numerical experiments of placing the electron randomly into the sample.}
\label{fig:OdKonformerow}
\end{figure}

\begin{figure}
\centering
\includegraphics[scale=0.3]{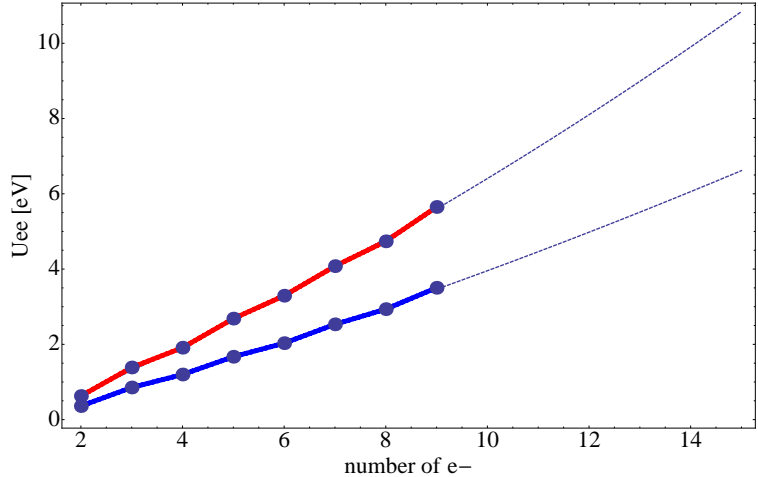}
\caption{The mean value of the interaction energy of one electron with other electrons in a function of electron number. The case with docosane as a medium (blue) is compared to the case without the medium (red). The dashed line denotes an extrapolation polynomial. The limit value of the ratio of the energies for these two cases equals about 2. The mean value is an average over 100 numerical experiments with randomly chosen positions of electrons.}
\label{fig:UeeOdIlosciElektronow}
\end{figure}

\begin{figure}
\centering
\includegraphics[scale=0.3]{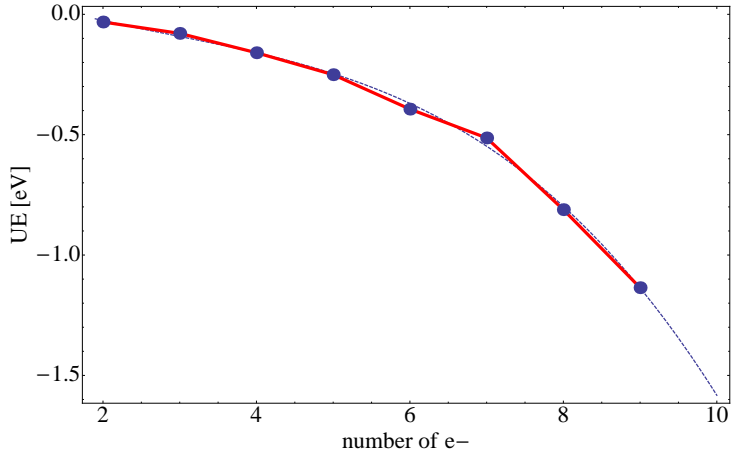}
\caption{The energy of the interaction of quasi-free electrons with the environment as a function of their number. Each point represents an average value of over 100 calculations with randomly chosen positions of electrons. The dashed line is an extrapolation polynomial.}
\label{fig:UEodIlosciElektronow}
\end{figure}

\begin{figure}
\centering
\includegraphics[scale=0.6]{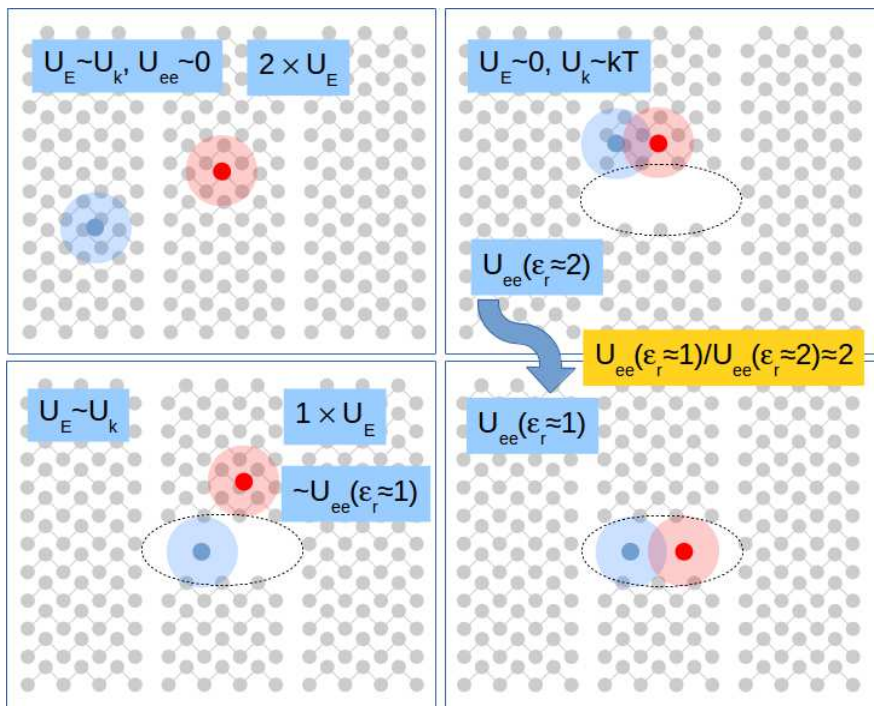}
\caption{Schematic presentation of typical electron and positron situations relative to the free volume. The notation is compatible with that used in the text. The cases are described in the text.}
\label{fig:Ps}
\end{figure}

Firstly, we calculated the energy of induced dipoles $U_E$ when one electron (positron) is inserted in the volume. The energy varies with the distance of the electron from the molecules. The dominant role is played by the nearest chunk here because the interaction potential changes rapidly with the distance. $U_E$ as a function of the nearest chunk distance from the electron is introduced in fig.\ref{fig:UE(r)}. For calculation of this energy, the electron's position was chosen randomly many times and the energy was calculated.
When the distance is less than about 2~$\AA$, the energy of the interaction of the electron with molecules starts to exceed thermal energy at room temperature (about 0.025~eV). This means that the electron that occupies this place does not travels in a free diffusion way any more. Possibly, it starts to move along the path of minimum energy and lands at a site that can be called a trap (our program does not contain an energy minimization procedure allowing us to follow the further evolution). If the distance is less than 1~$\AA$ the energy of the interaction starts to grow rapidly, even exceeding the value of 1~eV. These electrons can be called trapped (more detailed analysis of the energy of electron traps can be found in \cite{Pietrow15}).\\
We assume that the random choice of the position of an electron in a calculation represents the stage when an electron can move in a random walk (which is equivalent statistically to the diffusive motion \cite{Mehrer}). This kind of motion is realized in the period preceding the thermalization stage when the positron (electron) slows down by non-elastic scattering \cite{Mozumder}.\\
An interesting question is what fraction of electrons occupies a given distance from the nearest chunk. This problem can be solved numerically by collecting the set of cases of randomly chosen positions in a sample. The result of this numerical experiment is shown in figure~\ref{fig:RandomPlace}.
Most of the electrons occupy the distance about 4~$\AA$, which means (see fig.~\ref{fig:UE(r)}) that they move almost diffusively (the binding energy is lower than the thermal energy at room temperature). However, 27~\% of the population starts to adhere to the surrounding molecules after an initial diffusive jump; the energy of this bond is greater than the thermal one. 2~\% of the population is bound by the electrostatic interaction with an energy greater than 0.1~eV. These fractions can be called trapped electrons (for the spectrum of electron traps in alkanes see \cite{Pietrow13}).\\
The shape of the function depicted in fig.~\ref{fig:RandomPlace} resembles the Hertz's nearest-neighbor distribution function $H(r)$ of Poisson distributed points
\begin{equation}
H(r)=\rho \frac{dv(r)}{dr}e^{-\rho v(r)},
\end{equation}
where $v(r)$ is the volume of a spherical region of radius $r$, whereas $\rho$ is a number density~\cite{Torquato}. Obviously, in our case, the electron is not immersed in homogeneous surroundings. The shape of the function reflects the molecular arrangement in the alkane.\\
The interaction energy depends on the conformation. In a previous paper \cite{Pietrow15}, we showed that free volumes made by non-planar conformers form the deepest traps. Here, we investigate the influence of non-planar conformers on the non-trapped fraction of electrons. The results are presented in fig.~\ref{fig:OdKonformerow}. Each point in this figure denotes a median value of 300 numerical experiments of placing an electron into the sample in a random way. Most of them are located in places that allow quasi~-~free motion (fig.~\ref{fig:RandomPlace}) but in these cases the impact of the bond energy $U_E$ on the conformer composition is visible.
The numbers on the abscissa in fig.~\ref{fig:OdKonformerow} denote a probability of planar conformers (defined in sec.~\ref{Numerical assumptions}). Although the way of the numerical generation of non-planar conformers here is not precisely adequate to the actual construction of molecules (i.e. bond angles are non-adequate), the calculation shows a general rule when the number of bent molecules rises (diminishes the value on the abscissa). Namely, the bond energy $U_E$ diminishes with a rising probability of non-planar conformers. The result is easy to explain~--~the non-planar conformers can be related to a lesser degree of order. In this case, the induced dipoles, which interact with one another less constructively, make a cancellation of the field acting on a given chunk from an electron.\\
The density of non-planar conformers rises with the temperature in alkanes \cite{Maroncelli}. Our result coincides with the observation of emptying electron traps with temperature \cite{Pietrow06}. Indeed, the recent result shows that the energy of the bond with the surroundings diminishes with the number of non-planar conformers. If more of them appear with increasing temperature, the strength of the traps is lower.\\
If $n$ electrons are injected into the sample, they start to induce dipole momenta consecutively. The field generated by these dipoles moderates the interaction between the electrons. In this case, the electron -- electron interaction energy $U_{ee}$ is lower than the energy in the case of lack of the medium. Figure~\ref{fig:UeeOdIlosciElektronow} demonstrates the dependence of the interaction energy on the number of quasi-free electrons in the sample. Each point represents the median value of 100 results of energy calculation after random choice of the positions of $n$ electrons. The energy is calculated as a sum of electron -- electron interaction energy terms. In each case, the electric permittivity was calculated as indicated in sec.~\ref{Description of the model}.
$U_{ee}$ in the presence of docosane (blue in figure~\ref{fig:UeeOdIlosciElektronow}) was compared with the case when no medium is present (red). The dashed lines are polynomial fits of the series of points. In the limiting case ($n\rightarrow\infty$), the ratio of these energies tends to about 2. This means that the presence of alkane as a medium in which quasi-free electrons are immersed, moderates the interaction energy of electrons two times.\\
One can also calculate the interaction energy $U_E$ of $n$ electrons with induced dipoles. The electric field in place of each chunk is now a sum of the field vectors from these electrons. The result of these calculations is presented in fig.~\ref{fig:UEodIlosciElektronow}.
It is worth noting that $U_E$ is not a linear function of the number of electrons due to the collective interaction with the molecular medium. Here, the electron position was chosen from a rectangular region 40$\AA$ $\times$ 20$\AA$ $\times$ 160$\AA$ to facilitate overlapping the regions of electron interaction. It is expected that further addition of electrons (these calculations were not possible because of limitations of our hardware) finally makes a non-constructive induction of dipole momenta, and the energy will start to decrease.\\
Finally, consider the pair $e^+$ and $e^-$ in the medium. From the point of view of calculations here, both particles do induce dipole momenta equally well. Therefore, the positron can be regarded as confined to the environment or trapped. Because the positron annihilates in matter within a timescale comparable to a timescale known for other electromagnetic processes e.g. molecular vibrations (its lifetime is hundreds of picoseconds in the bulk), the frequency of oscillations of the chunks in the CDM model could be an important parameter here. For an unstable positron, the oscillation frequency of chunks could be inadequate to 'catch' the positron for a time compared to their lifetime. The characteristic frequency of dipoles in the CDM model is given by the formula $\omega=q^2/\sqrt{m\ \alpha}$, where $q$, $m$, $\alpha$ are a partially separated charge, mass, and polarizability of the chunk, respectively~\cite{Kwaadgras14}. Inserting numerical values, one can estimate the frequency of oscillations as $\nu\simeq$10$^{13}$~Hz (10 oscillations per picosecond). Thus, the feedback of the medium seems to be equally adequate to a positronium life scale.\\
An important problem in positron annihilation studies is to calculate the excess energy in the process of positronium atom formation in the free volume. This excess could inform whether the Ps formation is profitable in a particular case. Also, the consequence of deposition of this amount of energy is nontrivial (e.g. possible local melting). According to the model of positronium formation~\cite{Stepanov03}, just before Ps is formed in a free volume inside the medium, the $e^+$ -- $e^-$ pair migrates through the bulk as a loosely interacting pair called quasi-free Ps (qf-Ps). Consider the energy $U_E$ of the interaction of the pair with the bulk and mutual $e^+$~--~$e^-$ interaction energy $U_{ee}$ in four cases presented in figure~\ref{fig:Ps}.
If both the electron and the positron are placed randomly far from a free volume (upper left sub-figure), far from each other (an energy of mutual interaction is lower than 0.1~eV), and relatively far from the nearest chunks (most probable case), the energy of their interaction with induced dipoles of the bulk $U_E$ is of the order -0.1~eV. This means that a thermal energy at room temperature is large enough to make the configuration unstable.\\
If the particle pair assembles and settles the place near the free volume (upper right sub-figure), their interaction regions start to overlap and the induction of dipole momenta is cancelled, thus $U_E\simeq$0 (the particles loosen their traps). Both particles move as free particles with the thermal kinetic energy $U_k$. Their interaction energy $U_{ee}$ can be calculated as Coulomb interaction with the relative permittivity $\epsilon^{(r)}\simeq$2.\\
If the pair enters the free volume (lower right), $U_E$ can even increase because there could be more molecules in the region of the $e^+$ and $e^-$ interaction (the radius of the interaction was set empirically as 40$\AA$~--~see sec.~\ref{Numerical assumptions}). $U_E$ depends on the radius of the free volume but its value is negligible in relation to the interaction energy of the pair $U_{ee}$. In this case, $U_{ee}$ is Coulomb interaction energy for the relative permittivity $\epsilon^{(r)}\simeq$1. To summarize, the main energy excess when the pair transits to the free volume is related to the relative permittivity change. The energy changes twice.\\
It is also worth mentioning the case depicted in the lower left sub-figure of fig.~\ref{fig:Ps}. Because the relative permittivity here is almost 1, there is no excess energy when transition to Ps takes place.\\
The four cases described above relate to loosely bound $e^=$ particles that constitute most cases (fig.~\ref{fig:RandomPlace}) of particles moving in a random walk. However, in some percent of cases, the $e^=$ particles are not free. In this case, the bound energy should be taken into account and thus less advantage of Ps formation is expected.\\
\begin{figure}
\centering
\includegraphics[scale=0.6]{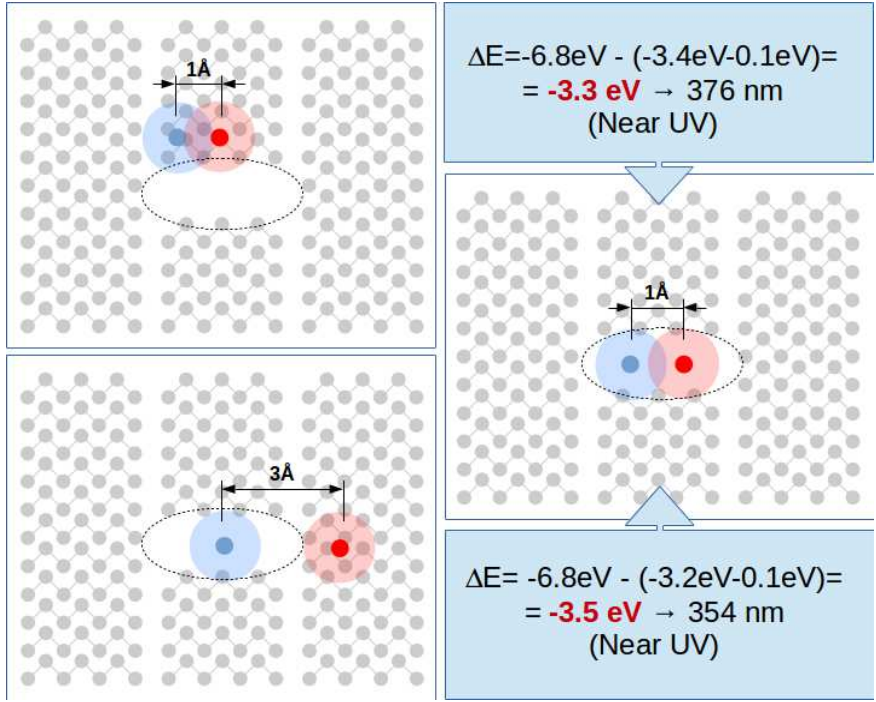}
\caption{Positron~--~electron pair near the free volume. The left column depicts two cases before Ps is formed. The right part~--~a schematic graph of Ps in a free volume. For the distances indicated in the sub-figures the excess energy was calculated for a transition from the left cases to the right one (blue fields).}
\label{fig:PsUV}
\end{figure}
Although the energy $U_E$ of coupling the pair to the medium is almost negligible in most cases (not trapped particles) in alkanes, $U_E$ can be regarded as perturbative contribution when the transition qf-Ps $\rightarrow$ Ps is considered. If one considers the transition described above and one set the numerical values of distances as realistic ones~--fig.~\ref{fig:PsUV}, the excess energy of transition to the positronium state is about 3~eV (here, the value for the initial state, i.e. the $e^+$ -- $e^-$ potential energy in a bulk plus the interaction energy of both particles with the bulk was subtracted from the final state energy, i.e. potential energy of Ps in the ground state in the free volume). The perturbation energy value is of an order of 0.1~eV. In the case of optical transition from the initial state to the final one, the spectrum is expected to be located in the near~-~UV (and visible) region and the variation of the photon energy due to the $U_E$ could characterize the properties of the medium. The measurement of the UV-Vis spectra in these cases could extend the skills of positron techniques. The transition from one of the states depicted on the left side of fig.~\ref{fig:PsUV} to that depicted on the right is characterized possibly by discrete levels of energy that could be calculated similarly to the hydrogenic model approach in spectroscopy of excitons \cite{Ibach09}. However, it cannot be presented now since it requires quantum mechanical modelling. It is a matter of experimental verification whether photonic Ps deexcitation exists at all. However, one should be aware of some experimental difficulties in photon deexcitation detection. These include not high efficiency of Ps formation signal in the case of typical low dose positron sources (the typical intensity of Ps formation in alkanes is tenths of percent), photon dispersion in matter and presence of photonic background from deexcitation of other atoms in the case of high energy positrons from the isotope source.\\
In most cases, the excess energy of Ps formation is deposited in phononic oscillations. The possibility of transition of this energy by collective vibrations should depend on the phononic spectrum of the medium. Furthermore, this energy should be transmitted by only few molecules surrounding the free volume. Possibly, non-collective vibrations and a local disorder (and modyfication of the volume) could be taken into account but it has not been reported yet. The Author is not able to consider the pros and cons of photonic and phononic transition details reliably but is seems that the competition of these two ways of releasing the energy would be medium dependent.
\section{Conclusions}
The Coupled Dipole Method allowed estimation of the energy of the electron and positron interaction in non-polar media. A systematic study of some typical cases was carried out and the energy balance during the positronium formation was calculated. The CDM method seems to be an adequate tool for the work function for electron (positron) estimation and for analysis of the energetic conditions in non-polar media, required in positron techniques. The model applied here is classical-mechanical one and could be designed as a 'first-order' approach. Given the classical formulas and point-particle approach, divergences of calculated parameters appear at the shortest distances. Nevertheless, the results justify the polarizability as an origin of the main source of interactions of the electron (positron) with alkanes. To get more accurate results, one should build a quantum model of the interaction where such effects as indistinguishability of electrons will be taken into account.\\
It was shown that some part of the electron (and positron) population is trapped in the medium and some interacts with the medium with an energy precluding their diffusive motion.\\
The calculations allow systematic investigations of energetic properties of a blob formed at the final stage of energetic positron track in the medium.\\
It was argued that if the excess energy of Ps formation is deposited in the photonic deexcitations, the optical spectra could be used as tools for exploration of medium properties, extending the positron techniques.
\section{Acknowledgements}
The author wants to thank Dr. Dominik Sza\l kowski (M. Curie-Sk\l odowska University, Lublin) for help with shell script works.\\
\\
This work was supported by the grant 2013/09/D/ST2/03712 of the National Science Center in Poland.
\bibliography{paper_Pietrow}{}
\bibliographystyle{unsrt}
\end{document}